\documentclass[aps,prb,reprint,superscriptaddress]{revtex4-2}
\usepackage{amsmath}
\usepackage{bm} 
\usepackage{amssymb}
\usepackage{titlesec}
\usepackage{color}
\usepackage{multirow}
\usepackage{url}
\usepackage[colorlinks,bookmarks=false,citecolor=red,linkcolor=blue,urlcolor=blue]{hyperref}
\usepackage{graphicx}
\usepackage{mathrsfs,bbm}
\usepackage{wrapfig}
\usepackage{physics}
\usepackage{enumerate}
\usepackage[version=3]{mhchem}
\usepackage{gensymb}
\usepackage{bm}
\usepackage{xcolor}

\def\chat{\hat{c}}
\def\nhat{\hat{n}}
\def\Hhat{\hat{H}}
\def\fhat{\hat{f}}

\begin{document}

	\title{Transmigration of Edge States with Interaction in Su-Schrieffer-Heeger Chain}
   	
	\author{Jyoti Bisht}
	\affiliation{School of Physical Sciences, Jawaharlal Nehru University, New Delhi 110067, India.}
	\author{Somenath Jalal}
	\affiliation{Department of Physics, Netaji Mahavidyalaya, Arambagh, Hooghly, West Bengal 712601, India.}
	\author{Brijesh Kumar}
	\email{bkumar@mail.jnu.ac.in}
	\affiliation{School of Physical Sciences, Jawaharlal Nehru University, New Delhi 110067, India.}
	\date{\today}
	
\begin{abstract}
The effect of Hubbard and Kondo interactions on the edge states in the half-filled Su-Schrieffer-Heeger chain of electrons is investigated by studying the behaviour of charge quasiparticles using Kumar representation and density matrix renormalization group method. For any finite dimerization of hopping, by increasing the Hubbard interaction, the edge states are found to transmigrate from the physical charge gap to a high energy gap through an intermediate phase without the edge states. The extent of this phase with no edge states shrinks smoothly upon increasing the dimerization. The transmigration of edge states from the charge gap to the high energy gap is also found to occur with Kondo interaction, but through an intermediate phase which itself changes from having no edge states for weak dimerization to having the edge states in the physical as well as the high energy gaps coexisting from moderate to strong dimerization. 
   \end{abstract}

\maketitle

\section{Introduction}

The Su-Schrieffer-Heeger (SSH) model was introduced historically to study solitons in conjugated polymers~\cite{su1979solitons,heeger1988solitons}. It describes tight-binding electrons in a half-filled one-dimensional lattice with dimerized hopping due to Peierls distortion~\cite{Peierls.book}. It is the simplest prototype of a topological insulator with edge states in the bulk gap~\cite{Asboth2016}. The subject of topological insulators is fundamentally concerned with studying band-structure topology and consequent edge (surface) states with inherently non-interacting models. An understanding of the effects of electron correlation on the topological surface states is most desired. Various studies find the electron-electron interaction to have a detrimental effect on the topological surface states~\cite{Rachel_2018}. Here we look afresh at this problem for the SSH model. In particular, we study the behaviour of edge states in the half-filled SSH chain with two basic interactions, namely, the Hubbard and Kondo interactions. The goal is to find out how exactly the SSH edge states react to and evolve with these common electronic interactions.

The half-filled SSH-Hubbard chain with dimerized hopping and local repulsion presents a minimal setting for the topological and correlation effects to compete. It has been studied in a variety of ways~\cite{manmana2012sshH,yoshida2014sshH,wang2015sshH,pachos2016sshH,ye2016entanglement,barbiero2018sshH,Le2020sshH,valenti2021sssH}, but a clear picture of the edge state behaviour in the interaction-dimerization plane is still found wanting. We relook at this problem by investigating the properties of the charge quasiparticles through an approach based on Kumar representation~\cite{kumar2008canonical}. This representation has been used fruitfully in studying interacting electron problems~\cite{kumar2009infiniteU,ram2017klm,faye2018,essler2018,ram2019spam,bertini2022,kurlov2023,pushkar2023klm}. Here we use Kumar representation in conjunction with DMRG (density matrix renormalization group) to work out an insightful and detailed phase diagram describing the edge state behaviour of the half-filled SSH-Hubbard chain. We apply the same approach for the half-filled SSH-Kondo chain in which the electrons interact with the localized quantum spin-1/2's via antiferromagnetic Kondo interaction. Incidentally, not much seems to be known about the edge states in the SSH-Kondo chain despite an interest in topological Kondo insulators. 
 
We formulate the problem of charge quasiparticles for the half-filled SSH-Hubbard and SSH-Kondo models in Sec.~\ref{sec:model}. It is used to study the behaviour of edge states in the two models. The results of our calculations for the SSH-Hubbard chain are presented in Sec.~\ref{sec:ssh-H}. From these calculations, we identify three distinct phases in the interaction-dimerization plane. For a fixed dimerization, the weakly correlated phase has two edge states in the physical charge gap, but the strongly correlated phase realizes the edge states in a high energy gap (relevant to quarter or three-quarter filling). In between these two phases lies an intermediate phase with no edge states. The extent of this intermediate phase increases monotonously upon decreasing the degree of dimerization. Next, in Sec.~\ref{sec:ssh-K}, we present our findings for the half-filled SSH-Kondo chain with a richer phase diagram. Here too, in going from weak to strong interaction, the edge states transmigrate from the physical to the higher energy gap. But the intermediate phase in this case turns out to be more subtle. It supports no edge states only for weak dimerization. For moderate or strong dimerization, it realizes the edge states in the charge gap and in the high energy gap simultaneously. We conclude this work with a summary in Sec.~\ref{sec:sum}.

\section{\label{sec:model} Charge Dynamics of Interacting SSH Models}
The model Hamiltonians of the SSH-Hubbard and SSH-Kondo chains, denoted respectively as $\Hhat_1$ and $\Hhat_2$, are given below.   
\begin{subequations}
\begin{eqnarray}
		\Hhat_{1} &=& \Hhat_{0} + U\sum_{l=1}^L\left(\hat{n}^{ }_{l,\uparrow} - \frac{1}{2}\right)\left(\hat{n}^{ }_{l,\downarrow} - \frac{1}{2}\right) \label{eq:sshh1} \\
		\Hhat_{2} &=& \Hhat_{0} + \frac{J}{2}\sum_{l=1}^L \vec{S}^{ }_l \cdot \vec{\tau}^{ }_l \label{eq:sshk}
\end{eqnarray}  
\end{subequations}
Here $U$ and $J$ are the Hubbard and Kondo interactions respectively, whereas 
\begin{equation}
\Hhat_0 = -t\sum_{l=1}^{L-1}\sum_{s=\uparrow,\downarrow} \left[1+(-)^l\delta\right]\left(\chat^\dag_{l,s}\chat^{ }_{l+1,s} + {\rm h.c.}\right) \label{eq:H0}
\end{equation}
is the SSH model with nearest-neighbour hopping on a dimerized one-dimensional lattice of total $L$ sites, with $0<\delta<1$ as the parameter of Peierls dimerization.  The Pauli operators $\vec{\tau}^{ }_l$ in $\Hhat_2$ describe the spin-1/2 local moments interacting on every site $l$ with the electron spin $\vec{S}_l$ given by $S^z_l = (\nhat_{l,\uparrow}-\nhat_{l,\downarrow})/2$ and $S^+_l=\chat^\dag_{l,\uparrow}\chat^{ }_{l,\downarrow}$. Since $\Hhat_1$ and $\Hhat_2$ are particle-hole symmetric, the zero chemical potential sets the electron filling to half.

Electrons in Kumar representation~\cite{kumar2008canonical} are described canonically by spinless fermions and Pauli operators. On one-dimensional bipartite lattice, the electron operators in Kumar representation can be written as: $\chat^\dag_{l,\uparrow} = [\fhat^\dag_l + (-)^l\fhat^{ }_l]\sigma^+_l$ and $\chat^\dag_{l,\downarrow} = \frac{1}{2}\{[\fhat^\dag_l - (-)^l\fhat^{ }_l] - [\fhat^\dag_l + (-)^l\fhat^{ }_l]\sigma^z_l\}$, in terms of the spinless fermions $\fhat_l$, and Pauli operators $\sigma^z_l$ and $\sigma^\pm_l$. Through the sign factor $(-)^l$, we represent the electrons on odd and even numbered sites in two different but equivalent forms. The SSH model in this representation reads as:
\begin{equation}
\begin{split}
\Hhat_0 = & -\frac{t}{2}\sum_{l=1}^{L-1} \left[1+(-)^l \delta\right] \left\{\left(\fhat^\dagger_l \fhat^{ }_{l+1} + {\rm h.c.}\right)\left(1+\vec{\sigma}_l\cdot\vec{\sigma}_{l+1}\right) \right. \\
	& \left. + (-)^l \left(\fhat^\dagger_l \fhat^\dagger_{l+1} + {\rm h.c.}\right)\left(1-\vec{\sigma}_l\cdot\vec{\sigma}_{l+1}\right)  \right\}, 
	\end{split} \label{eq:H0kumar}
\end{equation}
and the SSH-Hubbard and SSH-Kondo models take the following new forms: $\Hhat_{1} = \Hhat_{0} - \frac{U}{2}\sum_{l=1}^L \fhat^\dag_l\fhat^{ }_l + \frac{U}{4}L$ and $\Hhat_{2} = \Hhat_{0} + \frac{J}{4}\sum_{l=1}^L \fhat^\dag_l\fhat^{ }_l \left(\vec{\sigma}^{ }_l \cdot \vec{\tau}^{ }_l\right) $. In Kumar representation, they become the models of spinless `charge'  coupled with `spins'. Self-consistent treatment of charge and spin dynamics is one natural way to make progress in this form, and it is known to work for the half-filled correlated insulators~\cite{ram2017klm,ram2019spam,pushkar2023klm}. Thus, we study the properties of charge excitations of the SSH-Hubbard and SSH-Kondo chains by the effective model of spinless fermions given below; it is obtained by replacing the spin dependent operators in $\Hhat_1$ and $\Hhat_2$ by their bulk expectation values.
\begin{equation}
\begin{split}
\Hhat_c =& -\frac{t}{2}\sum_{l=1}^{L-1} \left[1+(-)^l \delta\right] \Big\{ \left[1+\rho^{ }_{1,(-)^l}\right] \fhat^\dagger_l \fhat^{ }_{l+1} + \\
& (-)^l \left[1-\rho^{ }_{1,(-)^l}\right] \fhat^\dagger_l \fhat^\dagger_{l+1} + {\rm h.c.}\Big\} + u \sum_{l=1}^L \fhat^\dag_l\fhat^{ }_l
\end{split} \label{eq:Hc}
\end{equation}
Here $u=-U/2$ for the SSH-Hubbard model and $J\rho_0/4$ for the SSH-Kondo model; $\rho_0$ is the average of $\langle \vec{\sigma}_l\cdot\vec{\tau}_l\rangle$ in the bulk. The $\rho_{1,(-)^l}$ is called $\rho_{1,-}$ for odd $l$'s 
 and $\rho_{1,+}$ for even $l$'s, which are obtained respectively by averaging 
 $\langle \vec{\sigma}_l\cdot\vec{\sigma}_{l+1}\rangle$ over the odd or even bonds in the bulk. 

This effective model of charge dynamics, $\Hhat_c$, needs $\rho_{1,\pm}$ and $\rho_0$ as inputs. Here we provide these inputs not approximately by self-consistency, but accurately by doing DMRG of the full SSH-Hubbard and SSH-Kondo chains. This hybrid approach, through $\Hhat_c$ with accurate input fields $\rho_{1,\pm}$ and $\rho_0$, enables us to study the precise nature of charge quasiparticles as canonical fermions. We perform DMRG calculations with our own code and also with ITensor~\cite{itensor}. 

Energy dispersion of the charge quasiparticles in the bulk can be calculated analytically for $\Hhat_c$ by assuming periodic boundary condition. By doing Fourier transformation, followed by Bogoliubov transformation, we can exactly diagonalize $\Hhat_c$ on a closed chain. The quasiparticle dispersions, thus derived, can be written as:
\begin{equation}
	E_{k,\pm} = \sqrt{u^2 + |\alpha_k|^2 + |\beta_k|^2 \pm 2\sqrt{u^2|\beta_k|^2 + \{\Re(\alpha^*_k\beta_k)\}^2}}
	\label{eq:Ekpm}
\end{equation}
where $\alpha_{k} =  t[(1+\delta)(1-\rho_{1+})+(1-\delta)(1-\rho_{1-})e^{i2k}]/2$ and $\beta_{k} =  t[(1+\delta)(1+\rho_{1+})+(1-\delta)(1+\rho_{1-})e^{i2k}]/2$ for $k\in[-\frac{\pi}{2},\frac{\pi}{2}]$. From this we get the bulk charge gap. 

The edge state behaviour of the SSH-Hubbard and SSH-Kondo chains is  investigated by solving Eq.~\eqref{eq:Hc} for the quasiparticle energies and wavefunctions by doing Bogoliubov diagonalization numerically 
on open chain. The findings from all these calculations are presented for the half-filled SSH-Hubbard chain in Sec.~\ref{sec:ssh-H}, and for the half-filled SSH-Kondo chain in Sec.~\ref{sec:ssh-K}.

\begin{figure}[b]
\begin{center}
\includegraphics[width=.8\columnwidth]{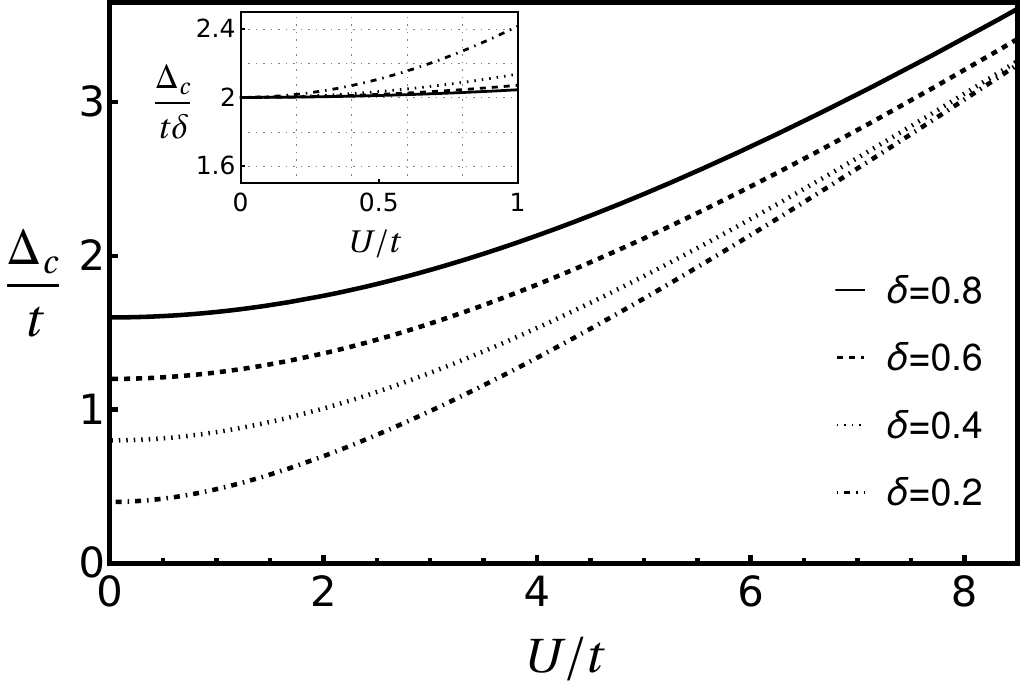}
\caption{Charge gap, $\Delta_c$, vs. Hubbard repulsion, $U$, for the half-filled SSH-Hubbard chain with different dimerization, $\delta$.}
\label{fig:gap-H}
\end{center}
\end{figure}
 
\section{\label{sec:ssh-H} Edge States in SSH-Hubbard Chain}
We investigate the behaviour of charge quasiparticles of the half-filled SSH-Hubbard chain for different strengths of dimerization, $0<\delta<1$, and Hubbard repulsion, $U>0$. We put $t=1$ in our calculations. For different values of $\delta$ and $U$, we first calculate the parameters $\rho_{1,\pm}$ for the SSH-Hubbard chain by DMRG. We get $0<\rho_{1,\pm}<-3$;  for strong values of $\delta$ and $U$, $\rho_{1,+}$ is found to be closer to -3, and $\rho_{1,-}$ closer to 0.  By putting the $\rho_{1,\pm}$ obtained from DMRG into $\Hhat_c$, we calculate the quasiparticle spectrum. It gives the charge gap, $\Delta_c$, presented in Fig.~\ref{fig:gap-H}. In the limit of small Hubbard repulsion, $\Delta_c$ correctly tends to $2t\delta$, the exact value for the non-interacting SSH chain at half-filling; see the inset of Fig.~\ref{fig:gap-H}. On the other hand, for large $U$, the charge gap tends to grow linearly with $U$, as expected. Clearly, this effective model of charge quasiparticles works for the entire range of $U$ and $\delta$. 

\begin{figure}[t]
\begin{center}
\includegraphics[width=\columnwidth]{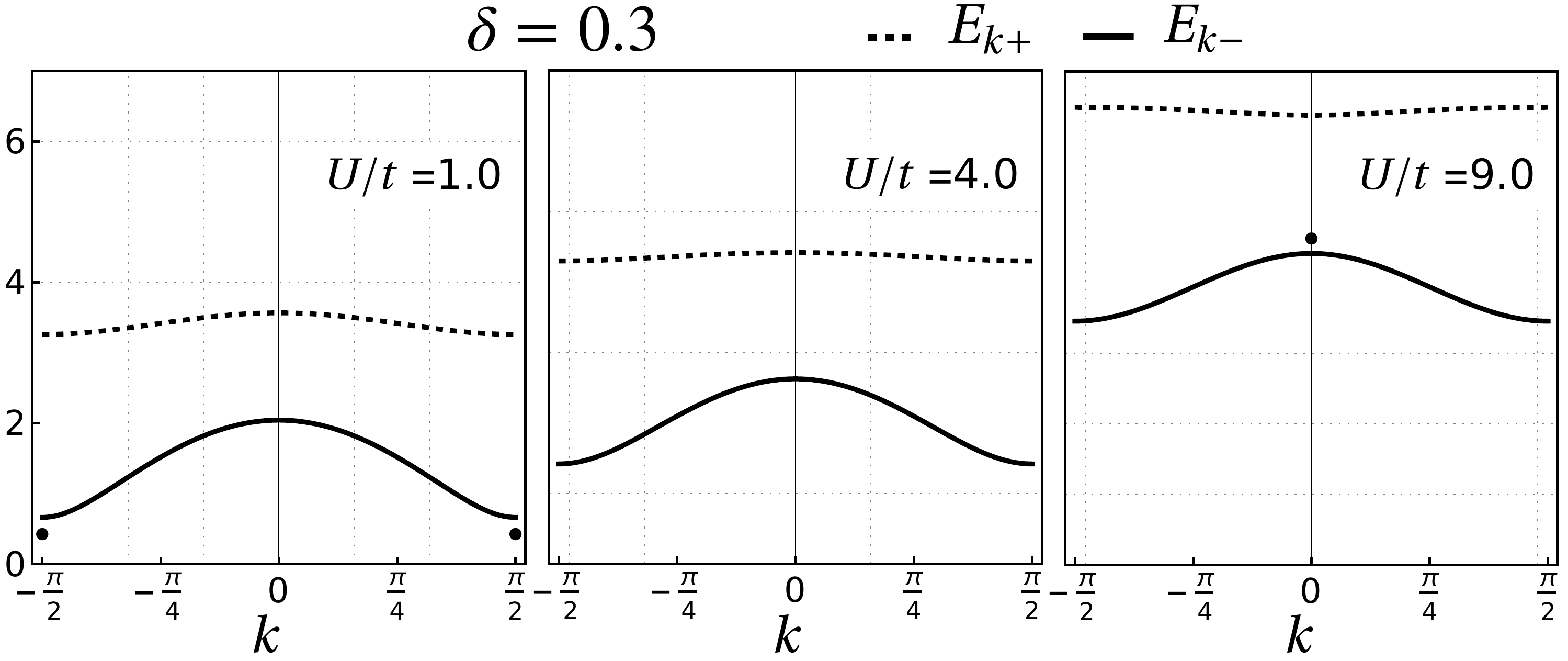}
\caption{Evolution of the quasiparticle dispersions, $E_{k,\pm}$ vs. $k$, with interaction for a given $\delta$ in the half-filled SSH-Hubbard chain. Note the transmigration of edge states (black dots) in the charge gap for small $U$ (first plot) to the high energy gap for large $U$ (third plot) through a stage with no edge states (second plot) for intermediate $U$.}
\label{fig:Ek-H}
\end{center}
\end{figure}

\begin{figure}[t]
\begin{center}
\includegraphics[width=\columnwidth]{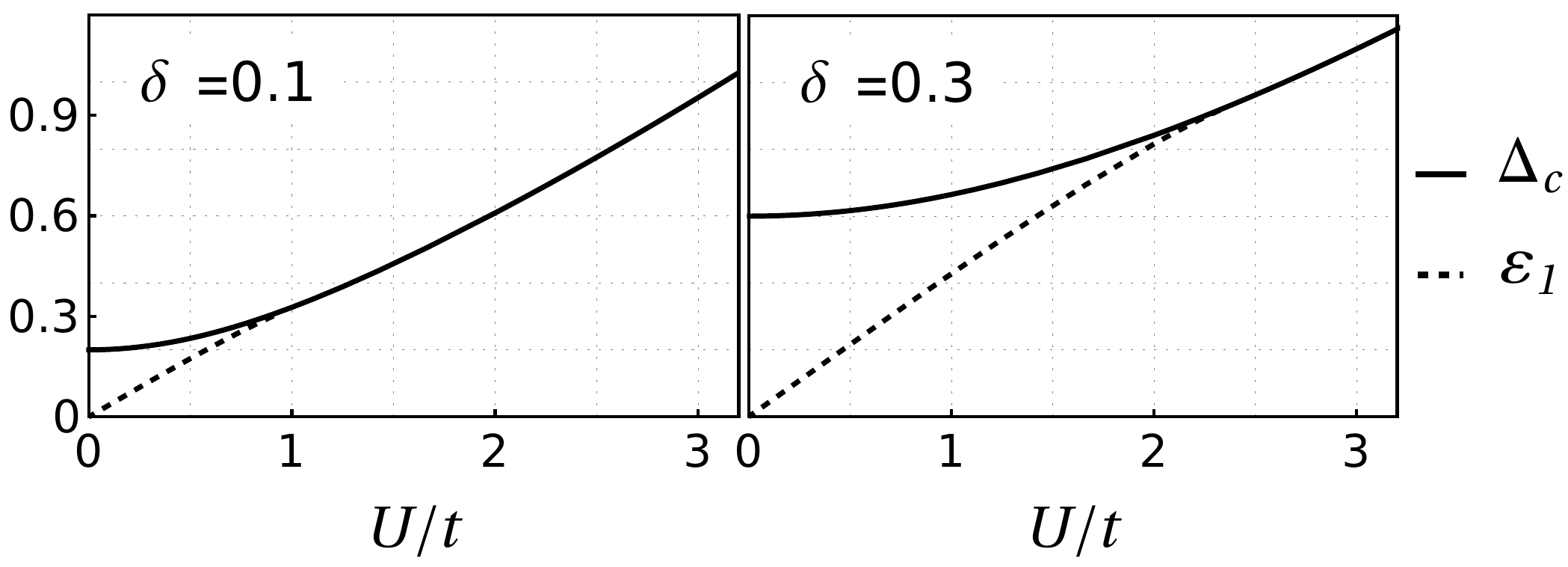}
\caption{The energy, $\varepsilon_1$, of the edge states in the bulk charge gap vs. $U$ for the half-filled SSH-Hubbard chain. The point where $\varepsilon_1$ equals $\Delta_c$ marks the critical point $U_{c,1}$ beyond which the edge states in the charge gap cease to exist.}
\label{fig:ES-energy}
\end{center}
\end{figure}

Let us look at the quasiparticle dispersions in more detail. See Fig.~\ref{fig:Ek-H}, where the bulk dispersions, $E_{k,\pm}$ given in Eq.~\ref{eq:Ekpm}, are plotted in the Brillouin zone, $ k \in [-\frac{\pi}{2},\frac{\pi}{2}] $, for three different values of $U$ for a fixed $\delta$. The minimum value of the lower energy dispersion $E_{k,-}$ is the physical charge gap, $\Delta_c$, presented in Fig.~\ref{fig:gap-H}. On open chain, for small $U$, we also get two edge states in the charge gap for any finite $\delta$. In the plot for $U=1.0$ and $\delta=0.3$ in Fig.~\ref{fig:Ek-H}, the two black dots at $k=\pm \frac{\pi}{2}$ mark these edge states. Notably, the edge states in the charge gap have a non-zero energy,  $\varepsilon_{1}$, that increases monotonously with $U$. Hence, upon increasing $U$, the edge states eventually overcome the charge gap at a critical interaction $U_{c,1}$, and are lost in the bulk; see Fig.~\ref{fig:ES-energy}. For a given $\delta$, we get a proportionately large $U_{c,1}$. However, this is not it. Upon increasing $U$ further, a second critical interaction $U_{c,2}$ is encountered beyond which the edge states reappear but in the high energy gap between $E_{k,-}$ and $E_{k,+}$, and not in the charge gap. In Fig.~\ref{fig:Ek-H} for $\delta=0.3$, the plot for $U=9$ shows the edge states in the high energy gap, whereas in the plot for $U=4$, the edges states are absent. 

\begin{figure}[t]
\begin{center}
\includegraphics[width=\columnwidth]{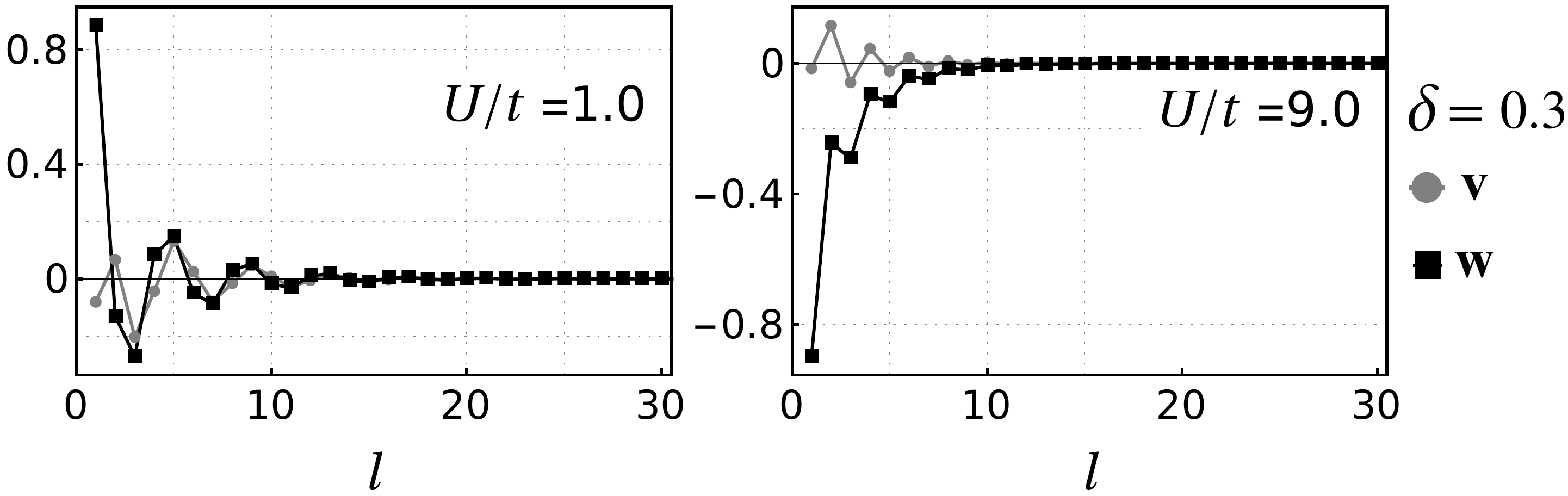}
\caption{The wavefunctions of the edge states for the half-filled SSH-Hubbard chain for $\delta=0.3$ for a small and a large value of $U$. In both cases, the wavefunctions decay with site label, $l$, but with an oscillatory modulation corresponding to wave-vector $k=\frac{\pi}{2}$ for small $U$ (i.e. for the edge state in the charge gap) and $k=0,\pi$ for large $U$ (i.e. for the edge state in the high energy gap).}
\label{fig:ES-wavefn-H}
\end{center}
\end{figure}

The wavefunctions of the edge states obtained on open chain by numerical Bogoliubov diagonalization of $\Hhat_c$ are found to be clearly localized at the opposite ends of the chain as shown in Fig.~\ref{fig:ES-wavefn-H}. A quasiparticle operator, $\hat{\eta}$, relates to the spinless fermions, $\fhat_l$, via Bogoliubov transformation:  $\hat{\eta} = \sum_{l=1}^L (v_l\fhat_l + w_l \fhat_l^\dag)$, where the vectors ${\bf v}$ and ${\bf w}$, with respective components $v_l$ and $w_l$, and normalization $|{\bf v}|^2+|{\bf w}|^2 =1$, carry the spatial profile of the quasiparticle. In Fig.~\ref{fig:ES-wavefn-H}, we plot the `wavefunctions' ${\bf v}$ and ${\bf w}$ of the edge states localized at the $l=1$ end of the chain for one small and one large value of $U$.  Notably, the spatial modulations present in these localized wavefunctions correspond to the wave-vector $k=\frac{\pi}{2}$ for the edge state in the charge gap (small $U$ case), and $k=0,\pi$ for the edge states the high energy gap (large $U$ case). This is why we have marked the two kinds of edge states in Fig.~\ref{fig:Ek-H} at $\pm \frac{\pi}{2}$ and $0$, respectively. 

\begin{figure}[b]
\begin{center}
\includegraphics[width=.78\columnwidth]{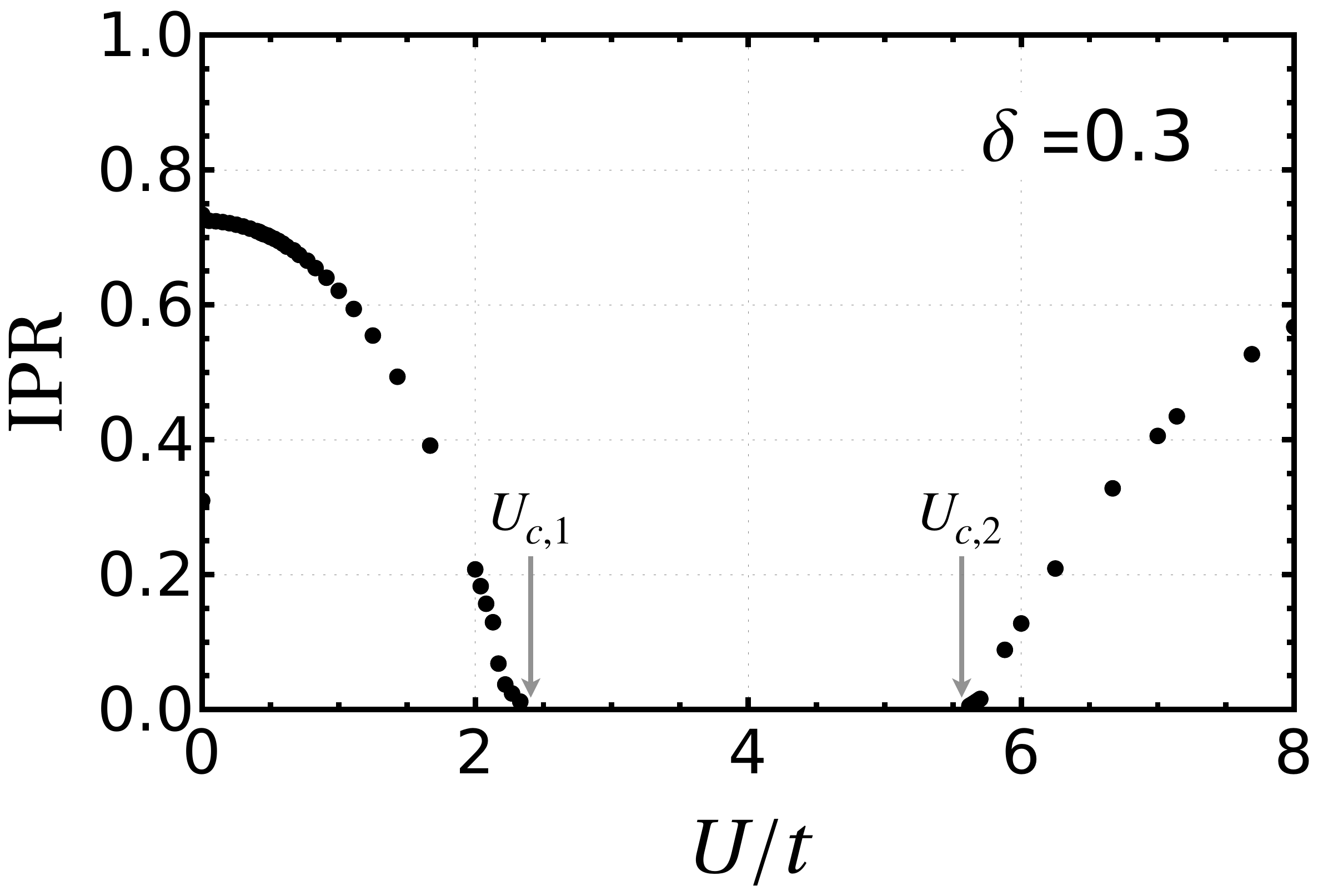}
\caption{Inverse participation ratio calculated as a function of $U/t$ for the eigenstates of $\Hhat_c$ on open chain. Non-zero IPR for small or large values of $U$ indicates the presence of edge states. Zero IPR in the middle for $U_{c,1}<U<U_{c,2}$ implies an absence of the edge states in the quasiparticle spectrum.}
\label{fig:IPR-H}
\end{center}
\end{figure}

Absence or presence of the localized states in the quasiparticle spectrum can also be tracked by inverse participation ratio (IPR), without having to look explicitly at the spatial profile of every wavefunction. The IPR is known to be zero for extended states, and non-zero for localized states. For a quasiparticle state given by vectors ${\bf v}$ and ${\bf w}$, it can be defined as: ${\rm IPR}= \sum_{l=1}^L \left(|v_l|^4 + |w_l|^4\right)$. We calculate the IPR as a function of $U/t$ for all the eigenstates of $\Hhat_c$ for a fixed $\delta$ on open chain. The data of such a calculation for $\delta=0.3$ is presented in Fig.~\ref{fig:IPR-H}. We find that, in the weakly as well as strongly correlated regimes of $U$, only two eigenstates of $\Hhat_c$ have non-zero IPR indicating clearly the presence of two edge states. However, in the intermediate range of interaction, $U_{c,1}<U<U_{c,2}$, the IPR is found to be zero for all the eigenstates implying no localized edge states. 

\begin{figure}[t]
	\centering
	\includegraphics[width=.88\columnwidth]{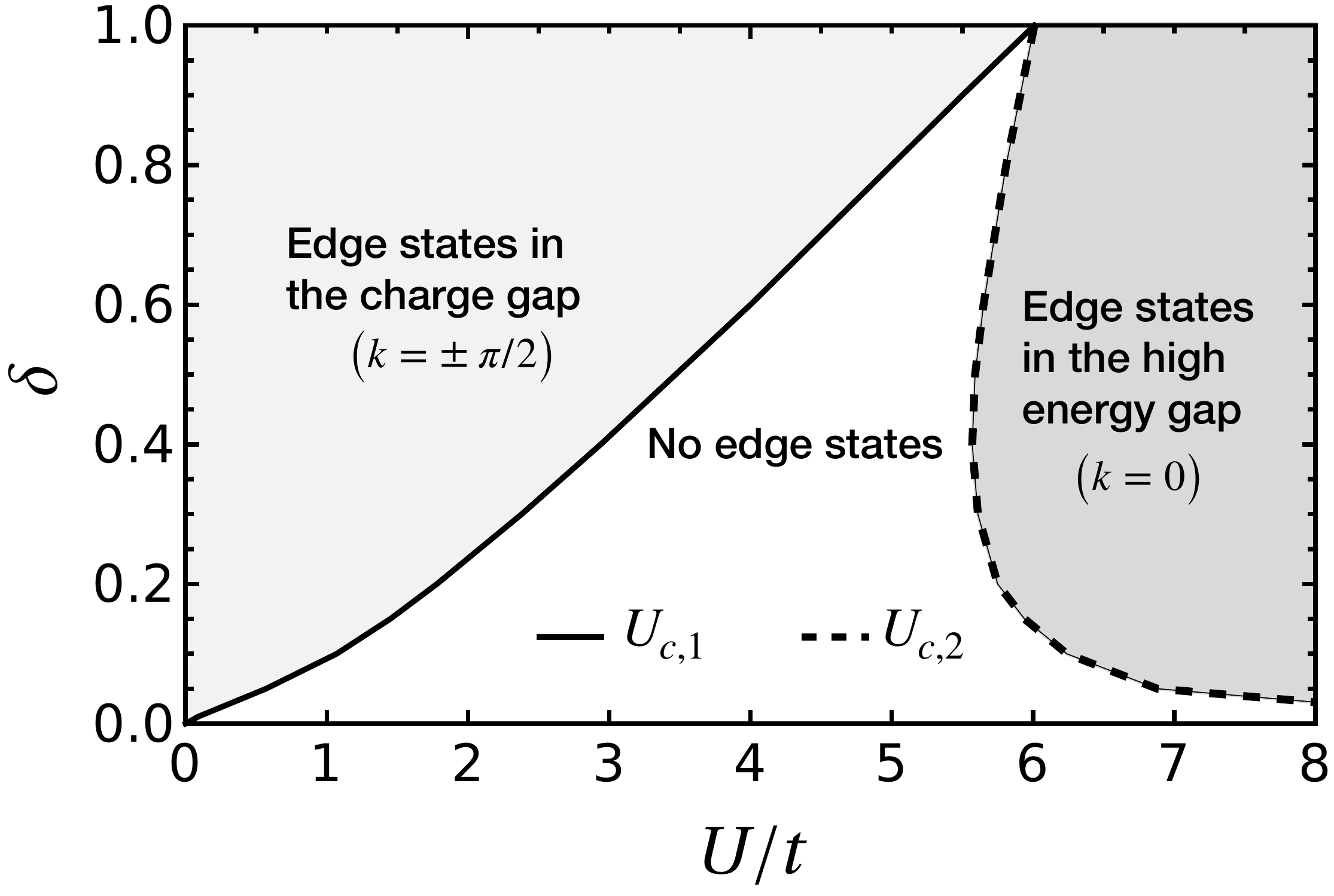}
	\caption{The phase diagram of the half-filled SSH-Hubbard chain based on the edge state behaviour of the charge quasiparticles. It has three phases demarcated by two boundaries, $U_{c,1}$ and $U_{c,2}$. In the weakly correlated phase for $U<U_{c,1}$, there exist two edge states in the charge gap at $k=\pm \pi/2$. In the intermediate phase for $U_{c,1} < U < U_{c,2}$, the edge states do not exist. In the strongly correlated phase for $U>U_{c,2}$, the edge states exist in the high energy gap at $k=0$.} 
		\label{fig:phase-diagram-H}
\end{figure}
 
These findings on the edge state behaviour of the half-filled SSH-Hubbard chain can be neatly summarized in the form of a phase diagram, Fig.~\ref{fig:phase-diagram-H}, in the interaction-dimerization plane. It has three phases, separated by two boundaries given by $U_{c,1}$ and $U_{c,2}$. The weakly correlated phase for $U<U_{c,1}$ realizes the edge states in the charge gap at $k=\frac{\pi}{2}$. There are no edge states in the intermediate phase given by $U_{c,1}<U<U_{c,2}$. In the strongly correlated phase for $U>U_{c,2}$, the edge states reappear, but in the high energy gap at $k=0$, and not in the charge gap. Thus,  for any given $\delta$, by increasing $U$, the edge states transmigrate from the physical charge gap to the high energy gap via a phase with no edge states. 

The edge states in the high energy gap are not relevant to the half-filled case as they do not lie in the charge gap. But they would be relevant for quarter or three-quarter fillings, for which this high energy gap would assume the role of physical charge gap. The dimerized Hubbard chain at quarter filling has been studied in the past~\cite{Penc1994}, but with no concerns for the edge states, except in one recent study~\cite{Le2020sshH}. Our study of the half-filled case finds the edge states relevant for quarter or three-quarter filling as the high energy edge states for strong correlations. 

In the absence of dimerization, i.e. $\delta=0$, the intermediate phase without edge states is of course the only phase for all values of the Hubbard interaction. For the extremely dimerized case of $\delta=1$, there is no intermediate phase, but only the two phases with edge states, which meet at $U/t=6$ exactly. In between these extremes, the width of the intermediate phase diminishes monotonously with increasing dimerization. Overall, this study presents a clear and interesting microscopic understanding of the edge states in the SSH-Hubbard chain. Next we investigate the SSH-Kondo chain for the effect of Kondo interaction on the edge states. 

\section{\label{sec:ssh-K} Edge States in SSH-Kondo Chain}
\begin{figure}[b]
\begin{center}
\includegraphics[width=.85\columnwidth]{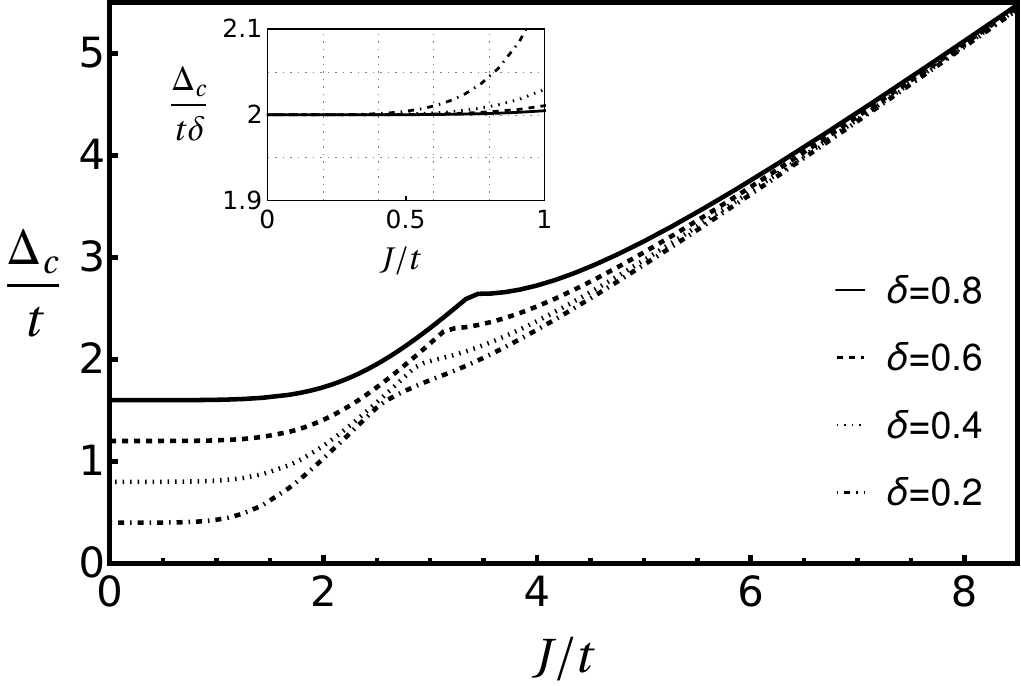}
\caption{Charge gap, $\Delta_c$, vs. Kondo interaction, $J$, for the half-filled SSH-Kondo chain with dimerization, $\delta$.}
\label{fig:gap-K}
\end{center}
\end{figure}

Here too we first do the DMRG calculation of $\rho_{1,\pm}$ and $\rho_0$ [defined below Eq.~\eqref{eq:Hc}] for $\Hhat_2$ with different values of $J>0$ and $\delta$ for $t=1$, and then use them as input parameters in $\Hhat_c$ to study the edge state behaviour of the charge quasiparticles of the half-filled SSH-Kondo chain. The charge gap, $\Delta_c$, obtained from this calculation is presented in Fig.~\ref{fig:gap-K}. For vanishingly small $J$, the charge gap correctly saturates to the value $2t\delta$ of the non-interacting SSH chain. For large $J$, it grows linearly with $J$. A kink in the charge gap at an intermediate $J$ signals a change to the Kondo singlet dominated regime for large $J$. 

\begin{figure*}[htbp]
\begin{center}
\includegraphics[width=\textwidth]{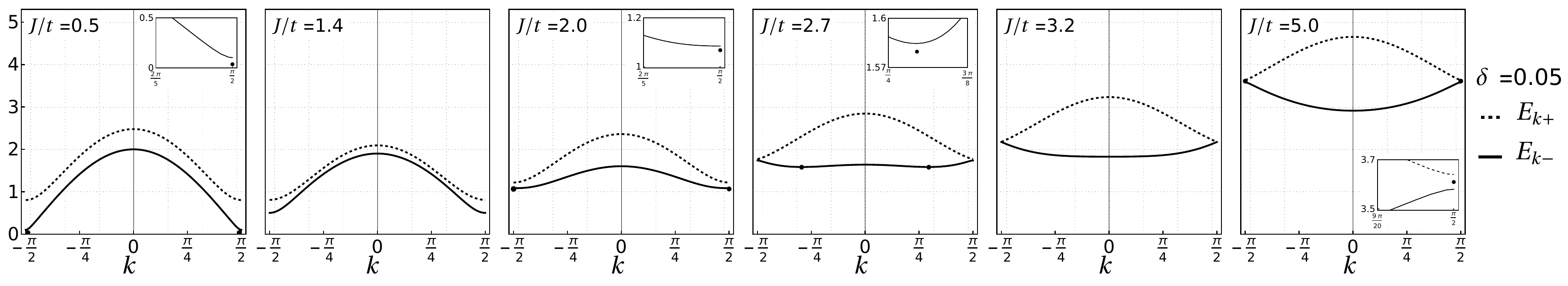}
\caption{Transmigration of the  edge states with $J$ in the half-filled SSH-Hubbard chain for $\delta=0.05$. At small $J$ the edge states (black dots) occur in the charge gap at $k=\pi/2$.  By increasing $J$, these edge states first disappear, then reappear and also shift gradually from $\pi/2$ towards 0 due to quasiparticle band inversion. By increasing $J$ beyond the inversion point $(J_i=3.11)$, the edge states again disappear, but then reappear in the high energy gap. Insets zoom in the details near the edge states.}
\label{fig:Ek-d05-K}
\end{center}
\end{figure*}

\begin{figure*}[htbp]
\begin{center}
\includegraphics[width=0.94\textwidth]{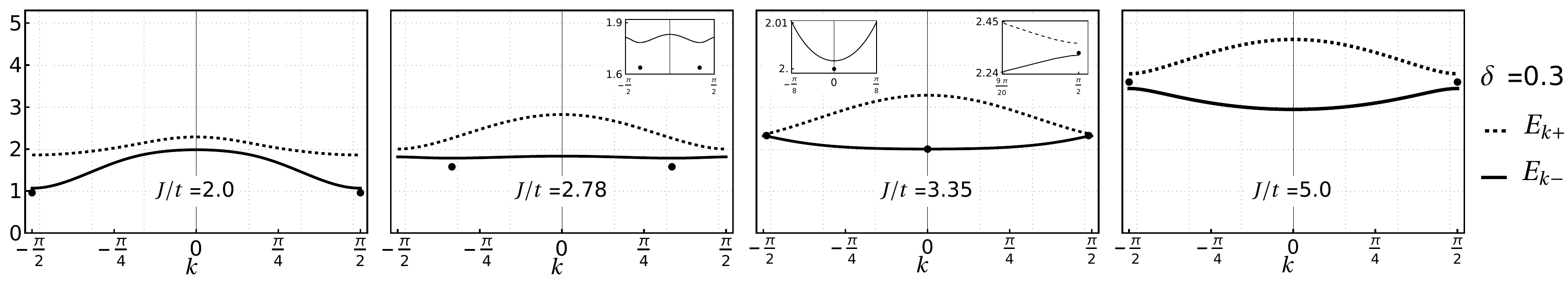}
\caption{Transmigration of the edge states with $J$ in the half-filled SSH-Hubbard chain for $\delta=0.3$. Here in a small range of intermediate $J$ beyond the inversion point, the edge states in the charge and the high energy gaps coexist; see the plot for $J=3.35$. Outside this range, the edge states exist only in the charge gap for small $J$ or in the high energy gap for large $J$.}
\label{fig:Ek-d3-K}
\end{center}
\end{figure*}

\begin{figure*}[htbp]
\begin{center}
\includegraphics[width=\textwidth]{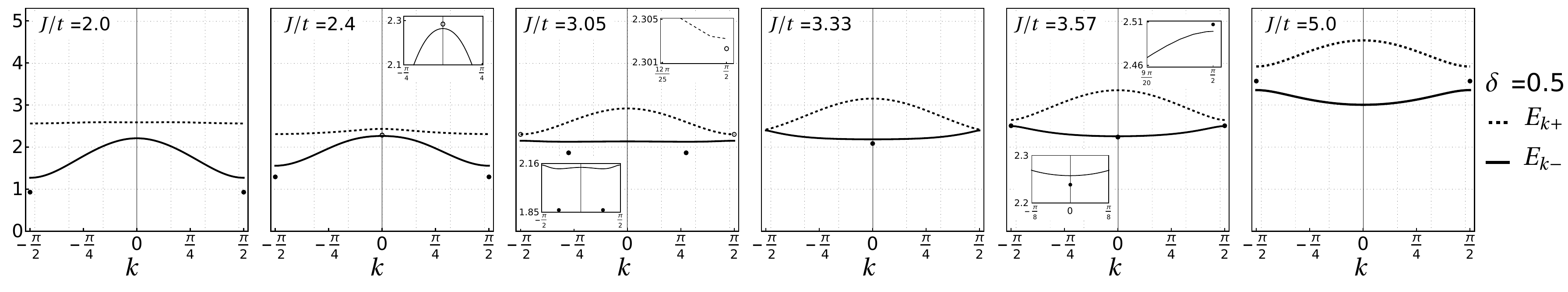}
\caption{Transmigration of the edge states with $J$ in the half-filled SSH-Kondo chain for $\delta=0.5$. Here the edge states in the charge gap and in the high energy gap coexist (as for $J=2.4$ and $3.57$) in small intervals of $J$ on both sides of the inversion point ($J_i=3.16$). Otherwise, the edge states occur only in the charge gap for small $J$, or in the high energy gap for large $J$.}
\label{fig:Ek-d5-K}
\end{center}
\end{figure*}

By computing the quasiparticle spectrum of $\Hhat_c$ on open chain, we get the edge states in the charge gap for small values of $J$. Upon increasing the Kondo interaction, these edge states also transmigrate from the charge gap to the high energy gap, but in a more complicated manner than what we saw for the SSH-Hubbard chain. Let us look closely at Figs.~\ref{fig:Ek-d05-K},~\ref{fig:Ek-d3-K} and~\ref{fig:Ek-d5-K} for the evolution of the quasiparticle spectrum with $J$ for three representative values of dimerization, $\delta=0.05$, 0.3 and 0.5, for the different manners in which this transmigration happens.

\begin{figure}[htbp]
\begin{center}
\includegraphics[width=\columnwidth]{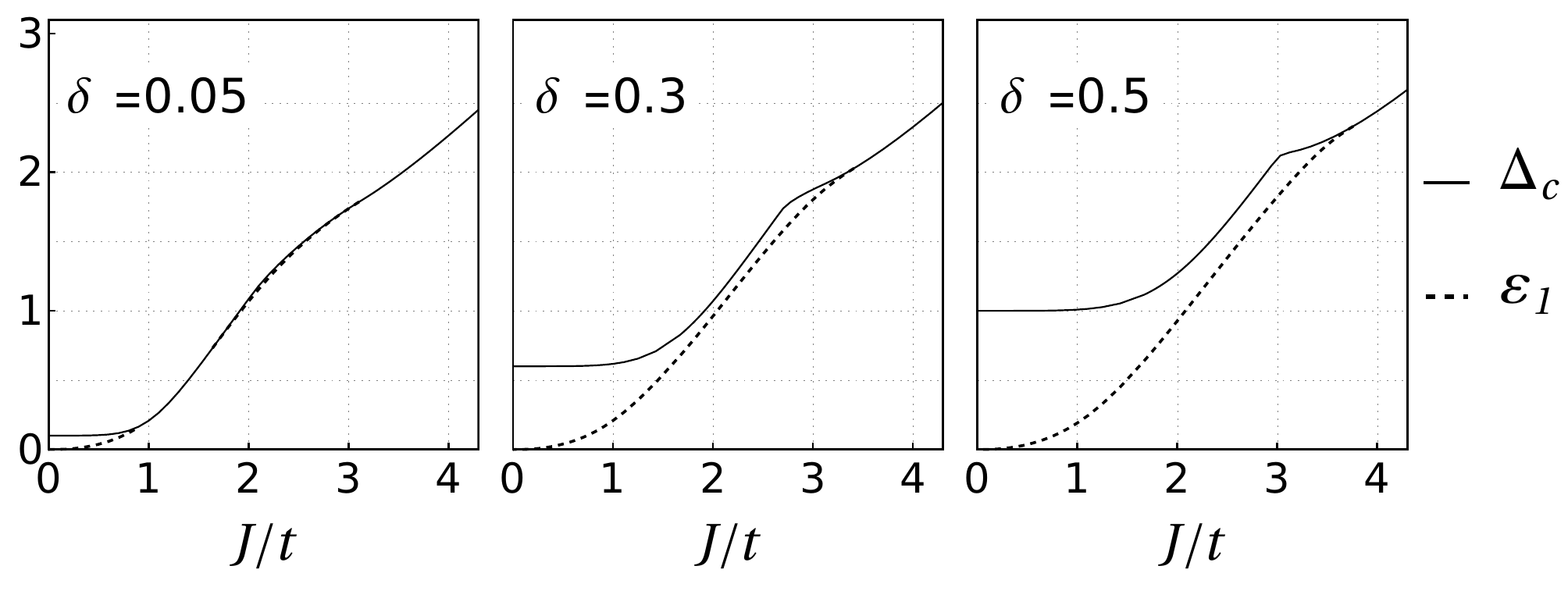}
\caption{The edge state energy, $\varepsilon_1$, and charge gap $\Delta_c$ vs. $J$ for the half-filled SSH-Kondo chain.}
\label{fig:ES-energy-K}
\end{center}
\end{figure}

\begin{figure}[htbp]
\begin{center}
\includegraphics[width=\columnwidth]{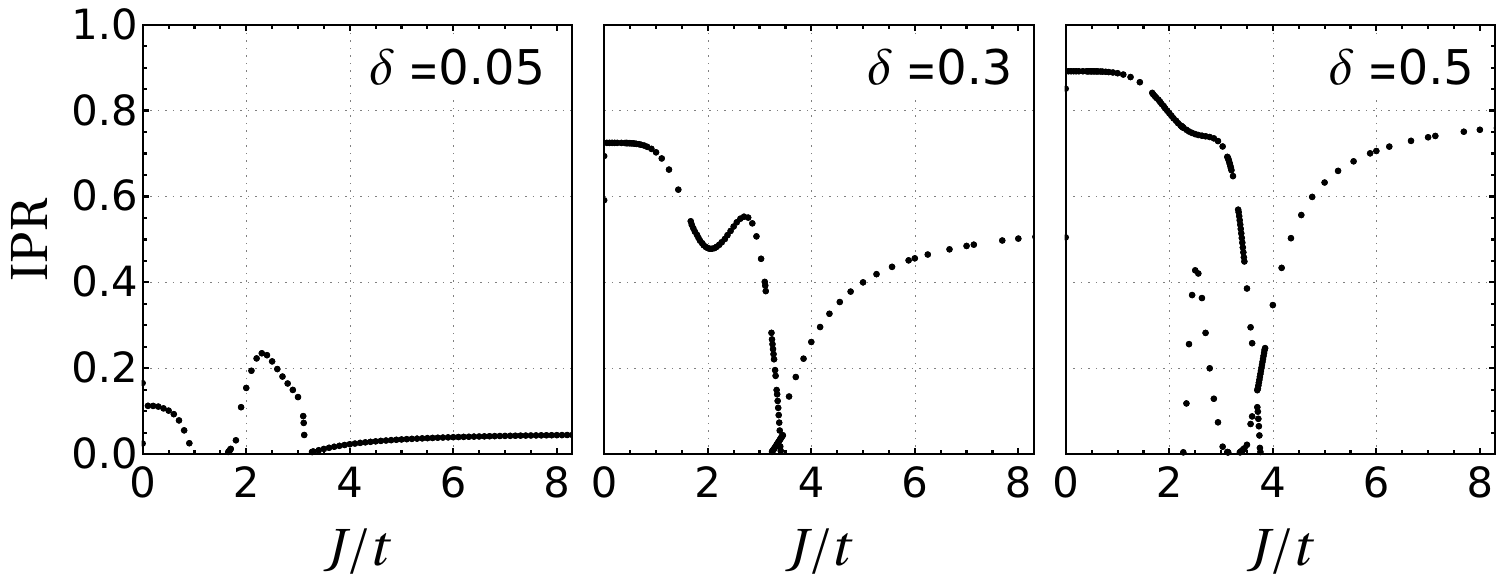}
\caption{Inverse participation ratio vs. $J$ for the eigenstates of $\Hhat_c$ on open chain for the half-filled SSH-Kondo model. Each data point represents two edges states. For $\delta=0.05$, the IPR is zero in two intervals of $J$ implying no edge states therein. For $\delta=0.3$, the IPR is always non-zero implying two edge states for small or large $J$, but four edge states in a small region of intermediate $J$. It is likewise for $\delta=0.5$, but with two such intermediate regions having four edge states.}
\label{fig:IPR-K}
\end{center}
\end{figure}

For $\delta=0.05$, the edge states for weak Kondo interaction lie in the charge gap. With an increase in $J$, these edge states disappear over a small range of $J$, and reappear again in the same charge gap. See the edge state energy $\varepsilon_1$ for $J=0.05$ in Fig.~\ref{fig:ES-energy-K}; after entering the bulk, it exits briefly and then reenters into the bulk. This behaviour is also clear from the data in Fig.~\ref{fig:IPR-K}, where the IPR for $\delta=0.05$ vanishes twice by increasing $J$. By increasing $J$, the dispersion $E_{k-}$ also undergoes inversion by gradually shifting the charge gap from $k=\pi/2$ to 0~\cite{ram2017klm,pushkar2023klm}. Beyond the inversion point, i.e. for $J>J_i=3.11$, $E_{k-}$ is always minimum at $k=0$, and the edge states in the charge gap disappear again. But for strong enough $J$, the edge states reappear in the high energy gap. The plots in Fig.~\ref{fig:Ek-d05-K} present for $\delta=0.05$ this sequence of changes in the quasiparticle spectrum. 

For $\delta=0.3$, the edge states exhibit a transmigration from the charge gap at small $J$ to the high energy gap for large $J$, but without ever completely disappearing in between. Instead, we find that for a small range of $J\gtrsim J_i$ (for $\delta=0.3$, $J_i=3.11$), the edge states in the charge gap coexist with the edge states in the high energy gap. See the plot for $J=3.35$ in Fig.~\ref{fig:Ek-d3-K}. Also see the plot for $\delta=0.3$ in Fig.~\ref{fig:IPR-K} wherein the IPR data from lower and higher $J$ sides overlap in small range of intermediate $J$. 

\begin{figure}[htbp]
\begin{center}
\includegraphics[width=\columnwidth]{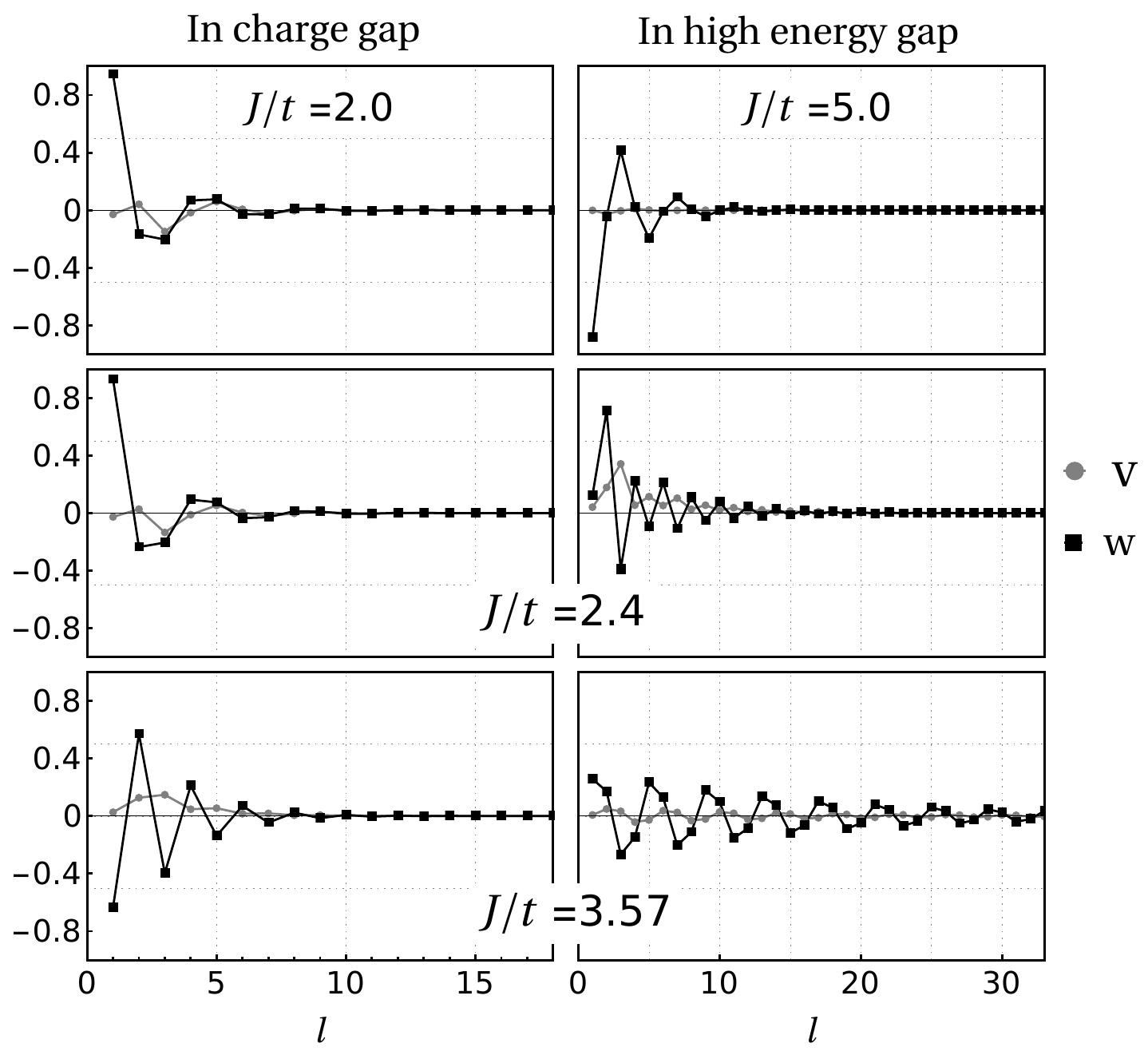}
\caption{Edge state wavefunctions for $\delta=0.5$ for a few different strengths of the Kondo interaction.}
\label{fig:wavefn-d5-K}
\end{center}
\end{figure}

The edge state behaviour of SSH-Kondo chain further changes as we increase the degree of dimerization. From the data for $\delta=0.5$ presented in Figs.~\ref{fig:Ek-d5-K} and~\ref{fig:IPR-K}, it is clear that the edge states in the charge gap and in the high energy gap coexist in two small regions, one on both sides of the inversion point, $J_i=3.16$. But there are two notable differences. Firstly, the high energy edge states below $J_i$ are found to peak at the second (or second-last) site, and not at the first (or the last) site. See in Fig.~\ref{fig:wavefn-d5-K} the wavefunctions for different values of $J$ in the charge and high energy gaps. Secondly, the coexistence region below $J_i$ is surrounded on both sides by a phase with edge states only in the charge gap. Whereas the coexistence region for $J>J_i$ is a region of overlap between two phases, one of which for lower values of $J$ has the edge states only in the charge gap, while the other for larger $J$'s realizes the edge states only in the high energy gap. 

\begin{figure}[htbp]
	\centering
	\includegraphics[width=.9\columnwidth]{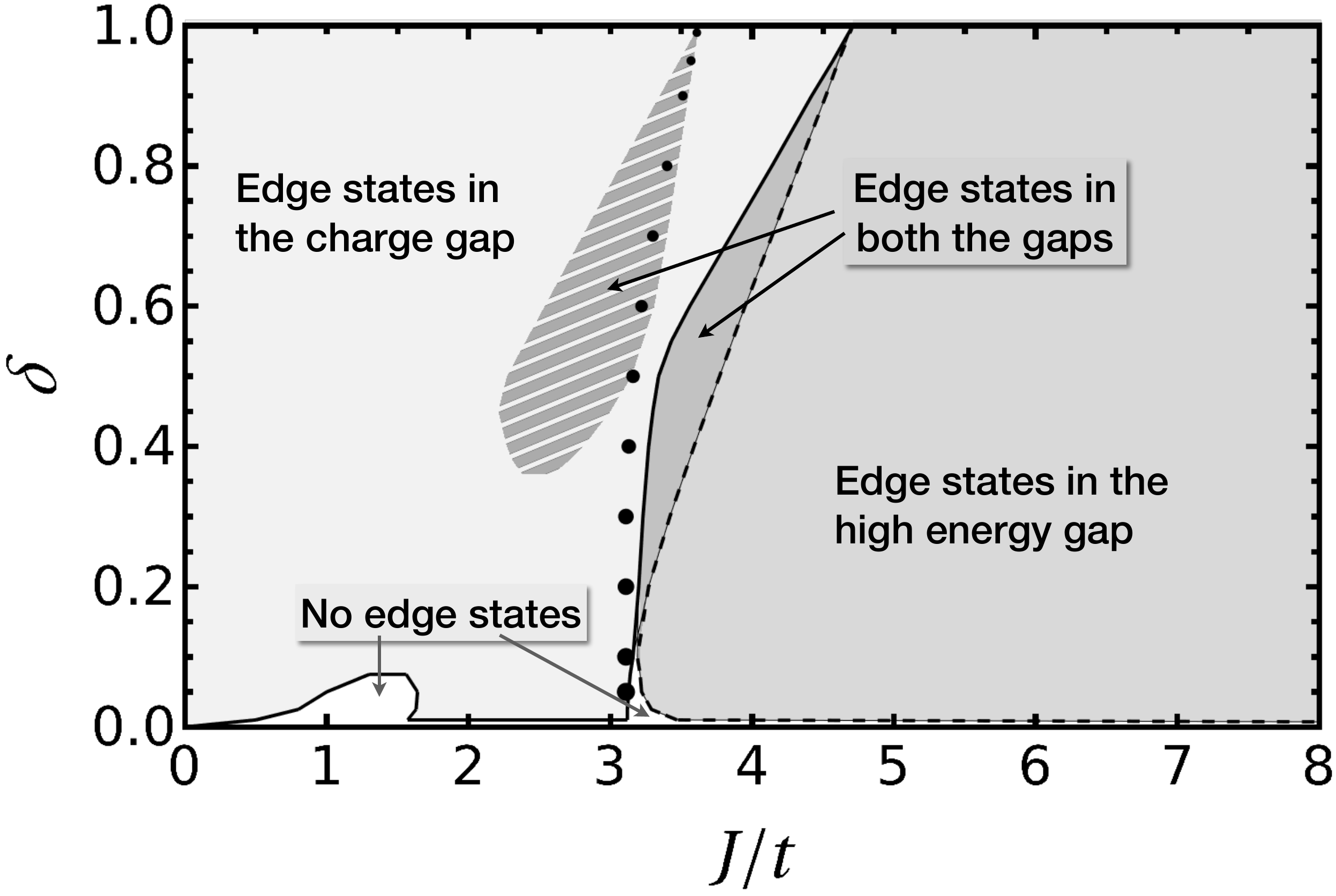}
	\caption{The phase diagram of the half-filled SSH-Kondo chain based on the edge state behaviour of the charge quasiparticles. It has five phases, as labelled. The line of black circles marks the inversion transition.}
		\label{fig:phase-diagram-K}
\end{figure}

The phase diagram in Fig.~\ref{fig:phase-diagram-K} sums up our findings on the edge state behaviour of the SSH-Kondo chain. It has four phases denoted by different background fillings. The white region for small $\delta$ is the phase without any edge states. The light gray region (extending from small to intermediate $J$) to enclosed by the solid line is the phase with two edge states in the charge gap. The gray region (extending from intermediate to large $J$) to bounded by the dashed line is the phase with two edge states in the high energy gap. The line of black circles mark the inversion transition. The hatched region to the left of the inversion line denotes a phase with two edge states in the charge gap and two in the high energy gap. The dark gray region to right of the inversion transition line denotes another phase with edge states coexisting in the charge gap and the high energy gap; it is sort of an intersection of the phases on itself left and right. 

The edge state behaviour of the SSH-Kondo chain can be grouped into three qualitative cases with respect to the degree of dimerization. For $0 < \delta \lesssim 0.1$, i.e. the weakly dimerized case, the edge states in the charge gap for small $J$ transmigrate to the high energy gap for large $J$, through an intermediate phase with no edge states. This case is somewhat like the SSH-Hubbard chain. For $0.1\lesssim \delta \lesssim 0.36$, i.e. the moderately dimerized case, the edge states in the charge gap transmigrate to the high energy gap with increasing $J$, but through an intermediate phase in which the edge states in both the gaps coexist. In the strongly dimerized case for $0.36 \lesssim \delta < 1$, the transmigration of the edge states from the charge gap to the high energy gap happens through two such intermediate phases with edge states in both the energy gaps. Clearly the SSH-Kondo chain exhibits a rich edge state behaviour in the interaction-dimerization plane. 

\section{\label{sec:sum} Summary}
This study of the half-filled SSH-Hubbard and SSH-Kondo chains presents an insightful understanding of the behaviour of the edge states with respect to interaction and dimerization. Its key finding is that the edge states which for weak correlations exist in the physical charge gap invariably transmigrate to the high energy gap for strong correlations. For the SSH-Hubbard chain, this transmigration of edges states with interaction happens through a simple intermediate phase with no edge states, in the same qualitative manner for different degrees of dimerization. For the SSH-Kondo chain, however, the intermediate phase changes with dimerization from having no edge states for weakly dimerized cases to realizing the edge states simultaneously in the physical as well as the high energy gap for moderate to strong dimerization. It would be interesting to explore this interaction driven transmigration of edge states in the wider context of topological insulators beyond the interacting SSH model. 

\acknowledgements{J.B. acknowledges DST (India) for INSPIRE fellowship, and thanks Arnav Pushkar for general discussions.}

\bibliography{manuscript}
	
\end{document}